\documentclass[%
  reprint,
  showpacs,preprintnumbers,
  amsmath,amssymb,
  aps,
  prb,
]{revtex4-1}

\usepackage{graphicx}
\usepackage{dcolumn}
\usepackage{bm}
\usepackage{hyperref}
\usepackage{color}
\usepackage{ulem}

\usepackage{times}
\usepackage{booktabs}
\usepackage{braket}

\begin{document}

\preprint{APS/123-QED}

\title{
Momentum-Dependent Spin Splitting by Collinear Antiferromagnetic Ordering
}

\author{Satoru Hayami$^1$, Yuki Yanagi$^2$, and Hiroaki Kusunose$^3$}
\affiliation{
$^1$Faculty of Science, Hokkaido University, Sapporo 060-0810, Japan \\
$^2$Center for Computational Materials Science, Institute for Materials Research, Tohoku University, Sendai, Miyagi, 950-8577, Japan \\
$^3$Department of Physics, Meiji University, Kawasaki 214-8571, Japan 
}
 
\begin{abstract}
We clarify the macroscopic symmetry and microscopic model-parameter conditions for emergence of spin-split electronic band structure in collinear antiferromagnets without atomic spin-orbit coupling. 
By using the microscopic multipole descriptions, we elucidate the fundamental degree of freedom in a cluster unit of an antiferromagnet giving rise to an effective spin-orbit interaction through the anisotropic kinetic motions of electrons. 
We show a correspondence of the ordering patterns and resultant momentum-dependent spin splitting for 32 crystallographic point groups after demonstrating two intuitive examples of four-sublattice pyrochlore and tetragonal systems. 
Our study unveils potential features of collinear antiferromagnets with considerably weak spin-orbit coupling in light-element materials and 3$d$ transition metal oxides, which can be utilized for a spin-current generation by electric (thermal) current and a magneto-striction effect. 
\end{abstract}
\maketitle

\section{Introduction}
The interplay between electronic degrees of freedom in solids is a source of fascinating physical phenomena in condensed matter physics. 
Among them, the atomic spin-orbit coupling (SOC) plays an essential role for leading to rich physics, such as the magnetoelectric effect~\cite{kimura2003magnetic,Fiebig0022-3727-38-8-R01,KhomskiiPhysics.2.20} and spin Hall effect~\cite{hirsch1999spin,Sinova_PhysRevLett.92.126603,bernevig2006quantum,sinova2015spin}.
The SOC-related physics has been extensively studied usually in materials containing heavier elements with the large atomic SOC~\cite{Herman_PhysRevLett.11.541}, where the novel electronic states and large physical responses have been discovered~\cite{levitov1985magnetoelectric,Fujimoto_PhysRevB.72.024515,Orenstein_PhysRevB.87.165110,Hayami_PhysRevB.90.024432,yoda2015current,Fu_PhysRevLett.115.026401,Zhong_PhysRevLett.116.077201}. 
Meanwhile, it tends to become cumbersome to control such phenomena flexibly because the atomic SOC is rooted in the complicated atomic orbital and chemical composition. 
In order to extend the scope of materials and explore further possibilities toward applications to next-generation electronics and spintronics devices~\cite{jungwirth2016antiferromagnetic,Baltz_RevModPhys.90.015005}, it is helpful to advocate another mechanism for SOC-related physics from a different viewpoint. 

In the present study, we discuss yet another intriguing interplay between the spin and orbital degrees of freedom that arises in the crystalline symmetry breaking through a spontaneous phase transition. 
Recently, several studies have shown that the electronic orders with lowering the lattice symmetry triggers unusual physical phenomena, such as the anomalous Hall effect in collinear antiferromagnets (AFMs)~\cite{vsmejkal2019crystal} and noncollinear AFMs~\cite{Suzuki_PhysRevB.95.094406,li2019quantum}, magneto-electric effects in the multipole orders~\cite{hitomi2016electric,Ishitobi_doi:10.7566/JPSJ.88.063708,hayami2016emergent,thole2018magnetoelectric}, spin-split Fermi surface in the electric toroidal order~\cite{Matteo_PhysRevB.96.115156,Hayami_PhysRevLett.122.147602}, the orbital Edelstein effect in the charge-density-wave state~\cite{massarelli2019orbital}, and the spin-current generation in organic AFMs~\cite{naka2019spin}. 
In particular, the last proposal is significant since the effective spin-orbit interaction is activated by the electronic order without the atomic SOC. 

Motivated by these studies, we push forward this issue in a more general framework to open another route of SOC physics in light-element materials, molecular organic metals, and 3$d$ transition metal oxides, whose atomic SOCs are negligibly small. 
We examine the symmetry conditions and microscopic parameters for the emergence of the effective spin-orbit interaction under the collinear AFMs without the atomic SOC. 
Our mechanism relies on neither an antisymmetric spin-orbit interaction as the Rashba metal~\cite{Dresselhaus_PhysRev.100.580,rashba1960properties} nor complex noncollinear and noncoplanar magnetic structures~\cite{Katsura_PhysRevLett.95.057205,tokura2010multiferroics}.  
By examining the microscopic tight-binding model based on multipole descriptions, we show that anisotropic kinetic motions of electrons in a collinear AFM gives rise to an effective spin-orbit interaction in momentum space.  
We demonstrate that the symmetric and anisotropic spin splittings arise in the four-sublattice pyrochlore and tetragonal systems as intuitive examples. 
Then, we show a systematic classification of the spin splitting in terms of specific AFM ordering patterns under 32 point groups, which provides a reference to explore physical phenomena driven by the spin-split band structures, such as a spin-current generation by electric (thermal) current and a uniform magnetization by a strain field, which is so-called a piezomagnetic effect.

The organization of this paper is as follows. 
In Sec.~\ref{Sec:Symmetry condition}, we show the symmetry conditions of the spin-split electronic structure with respect to the space-time inversion symmetry and any symmetric spin-split band dispersions are expressed as the even-parity electric multipole in momentum space. 
In Sec.~\ref{sec: Spin Splittings in tight-binding models}, we present microscopic ingredients to induce the spin-split band structures  in two tight-binding models on the basis of multipoles. 
In Sec.~\ref{sec:Physical phenomena by the spin splitting}, we discuss physical phenomena induced by the anisotropic spin-split band structure. 
Section~\ref{sec:Summary} is devoted to the summary. 
In Appendix~\ref{sec:Absence of spin splitting in a bipartite system}, we show that there is no spin-split band structure in a bipartite system due to the chiral symmetry. 
In Appendix~\ref{sec:Effective spin-orbit coupling for different phase conventions}, we derive the effective spin-orbit interaction in the presence of the collinear antiferromagnetic orderings. 
In Appendix~\ref{sec:Higher-order momentum-dependent spin splittings under 32 point groups}, we summarize types of spin-split band structures up to the sixth order in the wave number under 32 point groups. 

\section{Symmetry conditions}
\label{Sec:Symmetry condition}
\begin{figure}[t!]
\begin{center}
\includegraphics[width=1.0 \hsize]{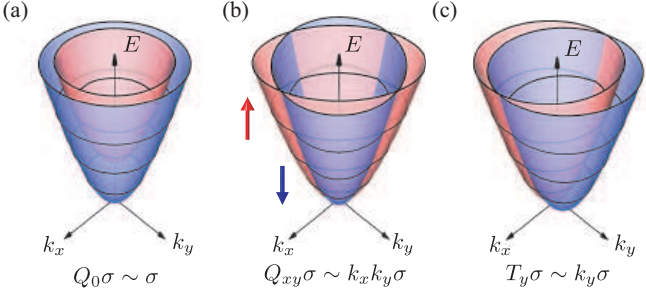} 
\caption{
\label{Fig:spinsplit_cm}
Schematic pictures of the spin splitting in collinear AFMs: (a) uniform ($\varepsilon_\sigma(\bm{k})\sim \sigma$), 
(b) anisotropic symmetric ($\varepsilon_\sigma(\bm{k})\sim k_x k_y\sigma$), and  (c) antisymmetric ($\varepsilon_\sigma(\bm{k})\sim k_y \sigma$) types. 
Symmetric modulations are characterized by the coupling between the spin $\sigma$ and even-parity (a) electric monopole $Q_0$ and (b) electric quadrupole $Q_{xy}$, while an antisymmetric one is due to odd-parity (c) magnetic toroidal dipole $T_y$. 
The red (blue) dispersions show the bands polarized with up (down) spins. 
}
\end{center}
\end{figure}

Let us start from the symmetry conditions of the spin-split electronic 
band structures in terms of the space-time inversion symmetry.
The general band dispersions $\varepsilon_{\sigma}(\bm{k})$ with the wave number $\bm{k}$ and the spin $\sigma$ are transformed with respect to the spatial inversion ($\mathcal{P}$) and time-reversal ($\mathcal{T}$) operations as $\mathcal{P}\varepsilon_{\sigma}(\bm{k})=\varepsilon_{\sigma}(-\bm{k})$ and $\mathcal{T}\varepsilon_{\sigma}(\bm{k})=\varepsilon_{-\sigma}(-\bm{k})$, respectively. 
Thus, the spin splitting $\varepsilon_{\sigma}(\bm{k}) \neq \varepsilon_{-\sigma}(\bm{k})$ requires the breaking of the $\mathcal{PT}$ symmetry, namely, either $\mathcal{P}$ or $\mathcal{T}$ must be broken at least. 
The breaking of $\mathcal{T}$ leads to the coupling between the even function of $\bm{k}$ and $\sigma$, i.e., the symmetric spin splitting with respect to $\bm{k}$, while the breaking of $\mathcal{P}$ gives rise to the coupling between the odd function of $\bm{k}$ and $\sigma$, which results in an antisymmetric spin splitting in momentum space as seen in the Rashba and Dresselhaus spin-orbit interactions~\cite{Dresselhaus_PhysRev.100.580,rashba1960properties}. 
Figures~\ref{Fig:spinsplit_cm}(a) and (b) 
represent the schematic pictures of the lowest and second-lowest symmetric spin-split band structure in the form of $\sigma$ and $k_x k_y \sigma$, respectively, while the result by a linear coupling $k_y \sigma$ is shown in Fig.~\ref{Fig:spinsplit_cm}(c).   
Note that the antisymmetric spin splitting in (c) can be usually caused in the presence of the atomic SOC.

In this paper, we focus on the collinear AFM orderings where the atomic SOC is negligible~\cite{comment_SS2}.
Thanks to SU(2) symmetry in spin space, the symmetry $\mathcal{R}\mathcal{T}$ ensures $\varepsilon_{\sigma}(\bm{k}) = \varepsilon_{\sigma}(-\bm{k})$ where $\mathcal{R}$ is the spin rotation.
This means that only the symmetric spin splitting appears even in noncentrosymmetric crystals.
Then, the spin-split band dispersion is generally expressed as 
\begin{eqnarray}
\label{eq:band}
\varepsilon_\sigma(\bm{k}) =\sum_{\Gamma\gamma} X_{\Gamma\gamma} Q_{\Gamma\gamma} (\bm{k}) \sigma,
\end{eqnarray}
where $Q_{\Gamma\gamma}(\bm{k})$ is the even-parity electric multipole in momentum space with the irreducible representation (irrep.) $\Gamma$ and its component $\gamma$~\cite{hayami2018microscopic,Hayami_PhysRevB.98.165110,Watanabe_PhysRevB.98.245129}. 
$X_{\Gamma\gamma}$ is the conjugate field that will be activated by the collinear magnetic ordering.
Note that the odd function of $\bm{k}$ in $\varepsilon_{\sigma}(\bm{k})$ can be expressed by the magnetic toroidal multipoles $T_{\Gamma\gamma}(\bm{k})$ in general~\cite{Hayami_PhysRevB.98.165110}, which are irrelevant in the present argument.
 
The set of electric multipole $Q_{\Gamma\gamma}(\bm{k})$ expresses all types of symmetric spin-split band structure.
For instance, the Zeeman spin splitting is represented by the monopole $Q_0 (\bm{k})=1$ [Fig.~\ref{Fig:spinsplit_cm}(a)].
The conjugate field $X_{0}$ corresponds to the molecular field (MF) of the spontaneous ferromagnetic ordering or an external magnetic field. 
Meanwhile, the higher-order multipoles such as a quadrupole $Q_{xy}(\bm{k})\sim k_{x}k_{y}$ give rise to anisotropic ($\bm{k}$-dependent) and symmetric spin splitting when an AFM ordering activates the corresponding conjugate field such as $X_{xy}$ [Fig.~\ref{Fig:spinsplit_cm}(b)].

\section{Spin Splittings in tight-binding models}
\label{sec: Spin Splittings in tight-binding models}
From the above consideration in Sec.~\ref{Sec:Symmetry condition}, the essence for the $\bm{k}$-dependent spin splitting is how to activate the anisotropic conjugate field $X_{\Gamma\gamma}$. 
As will be shown in this section, it is easy to be realized by collinear AFM orderings in crystal structures with sublattice degrees of freedom, where an anisotropic distribution of the ordered moments on the sublattice generates an anisotropic conjugate field $X_{\Gamma\gamma}$.
In order to clearly demonstrate the key microscopic parameters to activate $X_{\Gamma\gamma}$, we examine two tight-binding models on specific lattice structures: the pyrochlore structure in Sec.~\ref{sec:Pyrochlore structure} and the tetragonal structure in Sec.~\ref{sec:Tetragonal structure. T}he discussion is generalized to arbitrary point groups in Sec.~\ref{sec:General classification}. 

\subsection{Pyrochlore structure}
\label{sec:Pyrochlore structure}

\begin{figure}[t!]
\begin{center}
\includegraphics[width=1.0 \hsize]{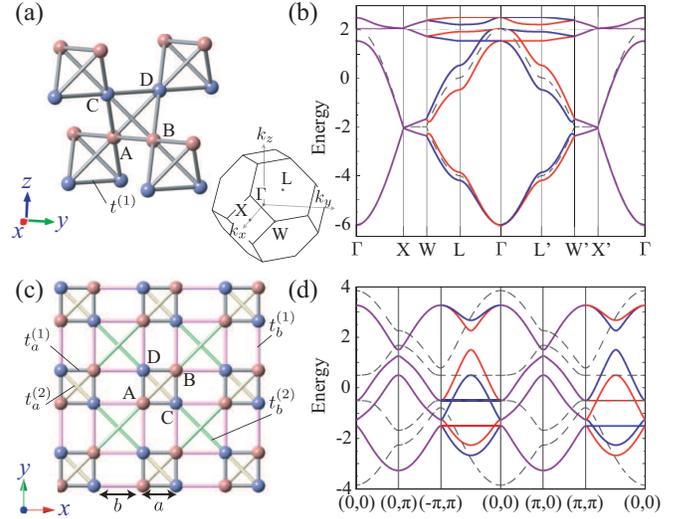} 
\caption{
\label{Fig:spinsplit}
Collinear AFM patterns (the red and blue spheres represent the opposite spin alignment) in (a) the pyrochlore structure and (c) the two-dimensional tetragonal structure, and the corresponding band structures in (b) and (d).
The hopping and MF parameters are given by $t^{(1)}=-1$ and $h_{xy}=0.5$ in (b), and by $t^{(1)}_a=1.1$, $t^{(2)}_b=1$, $t^{(1)}_{b}=t^{(2)}_{a}=0$, and $h_{xy}=0.5$ in (d).
The red (blue) lines show the up-(down-)spin bands. 
The dashed lines show spin-degenerate bands with $h_{xy}=0$ for (b), and with $t^{(1)}_b=0.8$ and $t^{(2)}_{b}=0$ for (d). 
The first Brillouin zone is shown in (b), where the prime symbols are related with $(k_x, k_y, k_z) \to (-k_x, k_y, -k_z)$.
}
\end{center}
\end{figure}

First, we analyze the three-dimensional pyrochlore structure with a unit of four-sublattice tetrahedron as shown in Fig.~\ref{Fig:spinsplit}(a). 
The positions of the four sublattice sites in the tetrahedron are defined by $\bm{r}_{\rm A} =(0, 0, 0)$, $\bm{r}_{\rm B} =(1/4, 1/4, 0)a$, $\bm{r}_{\rm C} =(1/4, 0, 1/4)a$, and $\bm{r}_{\rm D} =(0, 1/4, 1/4)a$, and we set $a=1$ as the unit of length. 
The single-orbital tight-binding model with the nearest-neighbor hopping $t^{(1)}$ is given by
\begin{eqnarray}
&\mathcal{H}_{0}&=\sum_{\bm{k}\sigma}\sum_{ij}c_{\bm{k}i\sigma}^{\dagger}H_{t}^{ij}c_{\bm{k}j\sigma}^{},
\cr&\quad&
H_{t}=2 t^{(1)}
\begin{pmatrix}
0 & c_{xy}^{+} & c_{zx}^{+} & c_{yz}^{+} \\
c_{xy}^{+} & 0 & c_{yz}^{-} & c_{zx}^{-} \\
c_{zx}^{+} & c_{yz}^{-} & 0 & c_{xy}^{-} \\
c_{yz}^{+} & c_{zx}^{-} & c_{xy}^{-} & 0
\end{pmatrix}
\quad
\begin{matrix} {\rm A} \\ {\rm B} \\ {\rm C} \\ {\rm D} \end{matrix}
\label{eq:HamtPyrochlore}
\end{eqnarray}
where $c^{\dagger}_{\bm{k}i\sigma}$ ($c_{\bm{k}i\sigma}^{}$) is the creation (annihilation) operator for wave vector $\bm{k}$, sublattice $i=$A-D, and spin $\sigma=\uparrow, \downarrow$, and $c_{\mu\nu}^{\pm}=\cos[(k_{\mu}\pm k_{\nu})a/4]$ for $\mu,\nu=x,y,z$.
Note that we consider the phase factor within a unit cell~\cite{comment_phase}. 
The effect of the uniform magnetic field or the collinear AFM ordering is described by the (MF) Hamiltonian $H_{t}\to H_{t}+H_{m}$,
\begin{eqnarray}
H_m=  (h_{0}+h_{yz}\rho_{z}\tau_{z}+h_{zx}\rho_{z}
+h_{xy}\tau_{z})\sigma,   
\label{eq:HamMF}
\end{eqnarray}
where the product of two Pauli matrices $\rho_\mu$ and $\tau_\nu$ represent the four sublattice degree of freedom, i.e., A-B and C-D, or (AB)-(CD) space, respectively. 
The first term represents uniform magnetic field, whereas the rest of terms express the MFs for three different collinear AFM patterns.

\begin{table}[t!]
\caption{
Multipoles classified by $T_{d}$ symmetry of the tetrahedron unit.
The lattice symmetry of the pyrochlore structure is indicated in the parenthesis.
The superscript represents the time-reversal parity. 
$c_{\mu \nu}=\cos (k_\mu a/4) \cos (k_\nu a/4)$, $s_{\mu \nu}=\sin (k_\mu a/4) \sin (k_\nu a/4)$ for $\mu,\nu=x,y,z$, and $c_{r}=c_{yz}+c_{zx}+c_{xy}$.
}
\label{tab_multipoles1}
\centering
\begin{tabular}{ccccc} \hline\hline
irrep. & type & $Q^{(0)}_{\Gamma\gamma}$ & $Q^{(1)}_{\Gamma\gamma}$ & $Q^{(1)}_{\Gamma\gamma}(\bm{k})$ \\ \hline
$A_{1}^{+}$ ($A_{1g}^{+}$) & $Q_{0}$ & $1$ & $\rho_{x}+\tau_{x}+\rho_{x}\tau_{x}$ & $\frac{2}{3}c_{r}t^{(1)}$ \\ \hline
$E^{+}$ ($E_{g}^{+}$) & $Q_{u}$ & & $\tau_{x}-2\rho_{x}+\rho_{x}\tau_{x}$ & $(\frac{1}{3}c_{r}-c_{xy})t^{(1)}$ \\
& $Q_{v}$ & & $\tau_{x}-\rho_{x}\tau_{x}$ & $(c_{zx}-c_{yz})t^{(1)}$ \\ \hline
$T_{2}^{+}$ ($T_{2g}^{+}$) & $Q_{yz}$ & $\rho_{z}\tau_{z}$ & $-\rho_{y}\tau_{y}$ & $-2s_{yz}t^{(1)}$ \\
& $Q_{zx}$ & $\rho_{z}$ & $\rho_{z}\tau_{x}$ & $-2s_{zx}t^{(1)}$ \\
& $Q_{xy}$ & $\tau_{z}$ & $\rho_{x}\tau_{z}$ & $-2s_{xy}t^{(1)}$ \\ \hline\hline
\end{tabular}
\end{table}

This model shows the $\bm{k}$-dependent spin splitting in the presence of any of $h_{yz}$, $h_{zx}$ or $h_{xy}$. 
Figure~\ref{Fig:spinsplit}(b) shows the band structure for $h_{xy}\ne 0$ for example. 
The origin of the spin splitting becomes transparent if one expresses the Hamiltonians in Eqs.~(\ref{eq:HamtPyrochlore}) and (\ref{eq:HamMF}) in terms of multipole language.
By introducing the multipoles as defined in Table~\ref{tab_multipoles1} according to symmetry operations of the tetrahedron unit, $T_{d}$ (the symmetry operations act only on the real space, not on the spin space), $H_{t}$ and $H_{m}$ are rewritten as
\begin{eqnarray}
&H_{t}&=
Q_{0}^{(1)}Q_{0}^{(1)}(\bm{k})+\left[Q_{u}^{(1)}Q_{u}^{(1)}(\bm{k})+Q_{v}^{(1)}Q_{v}^{(1)}(\bm{k})\right]
\cr&\quad&\quad
+\left[Q_{yz}^{(1)}Q_{yz}^{(1)}(\bm{k})+Q_{zx}^{(1)}Q_{zx}^{(1)}(\bm{k})+Q_{xy}^{(1)}Q_{xy}^{(1)}(\bm{k})\right],
\cr&
H_{m}&=\left[h_{0}Q_{0}^{(0)}+h_{yz}Q_{yz}^{(0)}+h_{zx}Q_{zx}^{(0)}+h_{xy}Q_{xy}^{(0)}\right]\sigma.
\label{eq:hammmul}
\end{eqnarray}
Since the same irreps. are coupled with each other, $h_{xy}$ term induces $Q_{xy}^{(1)}$ leading to the spin splitting via $Q_{xy}^{(1)}(\bm{k})$.
Thus, the spin splitting is characterized by $Q_{xy}^{(1)}(\bm{k})\sigma$ and its higher-order terms in $T_{2}$ irrep.
The spin splitting in cases of $h_{yz}\ne 0$ and/or $h_{zx}\ne 0$ is understood in a similar manner. 

\subsection{Tetragonal structure}
\label{sec:Tetragonal structure}
The spin splittings are ubiquitously found in other crystal structures irrespective of the spatial dimension and lattice symmetry. 
To demonstrate it, we further discuss the tetragonal crystal structure on the two-dimensional plane, as shown in Fig.~\ref{Fig:spinsplit}(c).
The positions of the four sublattice sites in the tetragonal unit with $C_{4v}$ symmetry are defined by 
$\bm{r}_{\rm A} =(-1/2, -1/2)a$, $\bm{r}_{\rm B} =(1/2, 1/2)a$, $\bm{r}_{\rm C} =(1/2, -1/2)a$, and $\bm{r}_{\rm D} =(-1/2, 1/2)a$ with $a+b=1$.
We consider the nearest- and next-nearest-neighbor hoppings (each contributions are indicated by the superscript) as shown in Fig.~\ref{Fig:spinsplit}(c), and the four-sublattice Hamiltonian is given in terms of multipoles defined in Table~\ref{tab_multipoles} as
\begin{eqnarray}
\label{eq:Hamt_123mp}
&H_{t}&=
\sum_{\eta=a,b}\biggl[
Q^{(1)}_{v}Q_{v}^{(1\eta)}(\bm{k})
+Q^{(2)}_{xy}Q_{xy}^{(2\eta)}(\bm{k})
\cr&\quad&
+\sum_{n=1,2} \biggl\{
Q^{(n)}_0Q_{0}^{(n\eta)}(\bm{k}) 
+\sum_{\zeta =x,y}T^{(n)}_\zeta T_{\zeta}^{(n\eta)}(\bm{k})
\biggr\}
\biggr],
\end{eqnarray}
and $H_{m}$ is given in the same form as Eq.~(\ref{eq:hammmul}). 
Here, the odd-parity magnetic toroidal dipoles $T_{\Gamma\gamma}(\bm{k})$ appear due to the lack of local inversion symmetry at each sublattice site in Fig.~\ref{Fig:spinsplit}(c).

\begin{table}[t!]
\caption{
Multipoles in two-dimensional tetragonal structure classified by $C_{4v}$.
The multipoles in the MF term are defined as $Q_0^{(0)}=1$, $Q_{yz}^{(0)}=-\rho_{z}$, $Q_{zx}^{(0)}=-\rho_{z}\tau_{z}$, $Q_{xy}^{(0)}=\tau_{z}$ in this case. 
$c_{\mu}^{\eta}=\cos k_{\mu}\eta$, $s_{\mu}^{\eta}=\sin k_{\mu}\eta$ for $\mu=x, y$ and $\eta=a,b$.
$p(a)=1$, $p(b)=-1$, and $\gamma_{\pm}=(\rho_{x}\tau_{x}\pm\tau_{x})/\sqrt{2}$.
}
\label{tab_multipoles}
\centering
\begin{tabular}{cccccc} \hline\hline
irrep. & type & $Q_{\Gamma\gamma}^{(1)}$ & $Q_{\Gamma\gamma}^{(2)}$ & $Q_{\Gamma\gamma}^{(1\eta)}(\bm{k})$ & $Q_{\Gamma\gamma}^{(2\eta)}(\bm{k})$ \\
& & $T_{\Gamma\gamma}^{(1)}$ & $T_{\Gamma\gamma}^{(2)}$ & $T_{\Gamma\gamma}^{(1\eta)}(\bm{k})$ & $T_{\Gamma\gamma}^{(2\eta)}(\bm{k})$\\ \hline
$A_{1}^{+}$ & $Q_{0}$ & $\gamma_{+}$ & $\rho_{x}$ & $\frac{1}{\sqrt{2}}(c_{x}^{\eta}+c_{y}^{\eta})t^{(1)}_{\eta}$ & $c_{x}^{\eta}c_{y}^{\eta}t^{(2)}_{\eta}$ \\ \hline
$B_{1}^{+}$ & $Q_{v}$ & $\gamma_{-}$ & & $-\frac{1}{\sqrt{2}}(c_{x}^{\eta}-c_{y}^{\eta})t^{(1)}_{\eta}$ & \\ \hline
$B_{2}^{+}$ & $Q_{xy}$  & & $\rho_{x}\tau_{z}$ & & $-s_{x}^{\eta}s_{y}^{\eta}t^{(2)}_{\eta}$ \\ \hline
$E^{+}$ & $Q_{yz}$ & $-\rho_z \tau_x$ & & & \\
& $Q_{zx}$ & $\rho_y \tau_y$ & & & \\ \hline
$E^{-}$ & $T_{x}$ & $-\rho_{z}\tau_{y}$ & $-\rho_{y}\tau_{z}$ & $p(\eta)s_{x}^{\eta}t^{(1)}_{\eta}$ & $p(\eta)s_{x}^{\eta}c_{y}^{\eta}t^{(2)}_{\eta}$ \\
& $T_{y}$ & $-\rho_{y}\tau_{x}$ & $-\rho_{y}$ & $p(\eta)s_{y}^{\eta}t^{(1)}_{\eta}$ & $p(\eta)s_{y}^{\eta}c_{x}^{\eta}t^{(2)}_{\eta}$ \\ \hline\hline
\end{tabular}
\end{table}

\begin{table*}[t!]
\caption{
Multipoles in the MF Hamiltonian $H_m$ and the hopping Hamiltonian $H_t$, spin splitting (SS) in the second order in $\bm{k}$, and symmetric spin current conductivity (SC) tensor under 32 point groups.
We also show types of spin-split band structures up to the sixth order in $\bm{k}$ in Appendix~\ref{sec:Higher-order momentum-dependent spin splittings under 32 point groups}. 
$k^2=k_x^2+k_y^2+k_z^2$.
We take the $x$ axis as the $C_{2}'$ rotation axis and take the $zx$ plane as the $\sigma_v$ mirror plane for $C_{3 {v}}$. 
The unlisted point groups are as follows: $T_{d}$ corresponds to $O$ when changing $T_1^- \to T_2^-$. 
$D_{2d}$ and $C_{4v}$ correspond to $D_4$ when changing $A_2^- \to B_2^-$ and $A_2^- \to B_1^-$, respectively. 
$S_{4}$ corresponds to $C_4$ when changing $A^- \to B^-$. 
$C_{6v}$ corresponds to $D_6$ when changing $A_2^- \to A_1^-$. 
$C_{3v}$ corresponds to $D_3$ when changing $A_2^- \to A_1^-$.
Note that the symmetry operations do not act on the spin space.
}
\label{tab_multipoles_table1}
\centering
\begin{tabular}{cccc|cc|cc|cc|cc|cccc|cc|c} \hline\hline
$H_m$  & $H_t$ & SS & SC tensor &
$O_{(h)}$ & $T_{(h)}$ &
$D_{4(h)}$ & $C_{4(h)}$ & 
$D_{2(h)}$ & $C_{2v}$ & $C_{2(h)}$ & $C_{s}$ &
$D_{6(h)}$ & $C_{6(h)}$ & $D_{3h}$ & $C_{3h}$ &
$D_{3(d)}$ & $C_{3(i)}$ & $C_{(i)}$ 
\\ \hline
$Q^{(0)}_{u}$ & $Q^{(n)}_{u}$ & $(3k_z^2-k^2)\sigma$ & $2\sigma^{\rm sc}_{zz}=-\sigma^{\rm sc}_{xx,yy}$
&
$E^+_{(g)}$ & $E^+_{(g)}$ & 
$A^+_{1(g)}$ & $A^+_{(g)}$ & 
$A^+_{(g)}$ & $A^+_{1}$ & $A^+_{(g)}$ & $A'^+$ 
& 
$A^+_{1(g)}$ & $A^+_{(g)}$ & $A'^+_1$ & $A'^+$ &
$A^+_{1(g)}$ & $A^+_{(g)}$  & $A^+_{(g)}$ 
\\
$Q^{(0)}_{v}$ & $Q^{(n)}_{v}$ & $(k_x^2-k_y^2)\sigma$ & $\sigma^{\rm sc}_{xx}=-\sigma^{\rm sc}_{yy}$ &
& &
$B^+_{1(g)}$ & $B^+_{(g)}$ & 
$A^+_{(g)}$ & $A^+_{1}$ & $A^+_{(g)}$ & $A'^+$ 
& 
$E^+_{2(g)}$ & $E^+_{2(g)}$ & $E'^+$ & $E'^+$ &
$E^+_{(g)}$ & $E^+_{(g)}$  & $A^+_{(g)}$ 
\\
$Q^{(0)}_{xy}$ & $Q^{(n)}_{xy}$ & $k_x k_y\sigma$ & $\sigma^{\rm sc}_{xy} $ &
$T^+_{2(g)}$ & $T^+_{(g)}$ & 
$B^+_{2(g)}$ & $B^+_{(g)}$ & 
$B^+_{1(g)}$ & $A^+_{2}$ & $A^+_{(g)}$ & $A'^+$ 
&
& & & &
 &  & $A^+_{(g)}$ 
\\ 
$Q^{(0)}_{yz}$ & $Q^{(n)}_{yz}$ & $k_y k_z\sigma$ & $\sigma^{\rm sc}_{yz}$ &
& &
$E^+_{(g)}$ & $E^+_{(g)}$ & 
$B^+_{3(g)}$ & $B^+_{2}$ & $B^+_{(g)}$ & $A''^+$ 
& 
$E^+_{1(g)}$ & $E^+_{1(g)}$ & $E''^+$ & $E''^+$ &
$E^+_{(g)}$ & $E^+_{(g)}$  & $A^+_{(g)}$ 
\\
$Q^{(0)}_{zx}$ & $Q^{(n)}_{zx}$ & $k_z k_x \sigma$ & $\sigma^{\rm sc}_{zx}$ &
& &
& & 
$B^+_{2(g)}$ & $B^+_{1}$ & $B^+_{(g)}$ & $A''^+$ 
&
& & & &  
 &  & $A^+_{(g)}$ 
\\
\hline
 & $T^{(n)}_{x}$ &  &
&
$T^-_{1(u)}$ & $T^-_{(u)}$ & 
$E^-_{(u)}$ & $E^-_{(u)}$ & 
$B^-_{3(u)}$ & $B^-_{1}$ & $B^-_{(u)}$ & $A'^-$ 
& 
$E^-_{1(u)}$ & $E^-_{1(u)}$ & $E'^-$ & $E'^-$ &
$E^-_{(u)}$ & $E^-_{(u)}$  & $A^-_{(u)}$ \\
 & $T^{(n)}_{y}$ &  &
&
 &  & 
 & & 
$B^-_{2(u)}$ & $B^-_{2}$ & $B^-_{(u)}$ & $A'^-$ 
& 
 & & & &
 &  & $A^-_{(u)}$ \\
 & $T^{(n)}_{z}$ &  &
&
 &  & 
$A^-_{2(u)}$ & $A^-_{(u)}$ & 
$B^-_{1(u)}$ & $A^-_{1}$ & $A^-_{(u)}$ & $A''^-$ 
& 
$A^-_{2(u)}$ & $A^-_{(u)}$ & $A''^-_2$ & $A''^-$ &
$A^-_{2(u)}$ & $A^-_{(u)}$  & $A^-_{(u)}$ \\
\hline\hline
\end{tabular}
\end{table*}

The AFM orderings can activate only in $B_{2}^{+}$ or $E^{+}$ irreps. as shown in $H_{m}$ in Eq.~(\ref{eq:hammmul}).
Moreover, the coupled multipoles in momentum space only exist in $B_{2}^{+}$ irrep. from the $B_{2}^{+}$ row in Table~\ref{tab_multipoles}.
This indicates that the spin-split band structure can be obtained only in the case of $h_{xy}\ne 0$ and $t_{\eta}^{(2)}\ne 0$, while there is no spin splitting under the AFM ordering in $E^{+}$ irrep, i.e. $h_{yz}\neq 0$ or $h_{zx}\neq 0$, due to the lack of the coupled multipoles in momentum space. 
Figure~\ref{Fig:spinsplit}(d) indeed shows the spin splitting for $t^{(1)}_a=1.1$, $t^{(2)}_b=1$, $h_{xy}=0.5$, and $t_{a}^{(2)}=t_{b}^{(1)}=0$, which is characterized by $Q_{xy}^{(2\eta)}(\bm{k})$, i.e., $\varepsilon_{\sigma}(\bm{k})\sim k_x k_y \sigma$. 
However, by taking finite $t^{(1)}_b=0.8$ instead of $t^{(2)}_b$, the spin splitting does not appear, as shown by the dashed lines in Fig.~\ref{Fig:spinsplit}(d) where each band is doubly degenerate.
This is because the additional chiral symmetry in a bipartite system at $t_{a}^{(2)}=t_{b}^{(2)}=0$ prohibits the activation of the relevant multipoles, as discussed in Appendix~\ref{sec:Absence of spin splitting in a bipartite system}.

These examples of the pyrochlore and tetragonal structures clearly show that new type of effective spin-orbit interaction is activated by the collinear AFM orderings in sublattice systems.
In the latter 
example of this section, the magnitude of the MF in Eq.~(\ref{eq:band}) leading to the spin splitting is evaluated as $X_{xy}\sim {\rm sgn}(t^{(2)}_b)(t^{(1)}_{a})^2/h_{xy}$ for small $t^{(1)}_{a}$ and $t^{(2)}_b$. 
This indicates a potential aspect of realizing the large spin splitting by combining hopping amplitude and a spontaneous magnetic ordering even without the atomic SOC.
The necessary condition for such a $\bm{k}$-dependent spin splitting is summarized as follows: the hopping matrix $Q_{\Gamma\gamma}^{(n)}(\bm{k})$ ($n\ge1$) or its higher-order multiplication such as $T_{x}^{(1\eta)}(\bm{k})T_{y}^{(1\eta)}(\bm{k})$ belongs to the same irrep. $\Gamma$ of the MF multipole $Q_{\Gamma\gamma}^{(0)}$ generated in the ordered state.
The direct effective coupling between the bond multipoles in $H_t$ and the MF multipoles in $H_m$ is shown in Appendix~\ref{sec:Effective spin-orbit coupling for different phase conventions}. 

\begin{figure*}[t!]
\begin{center}
\includegraphics[width=1.0 \hsize]{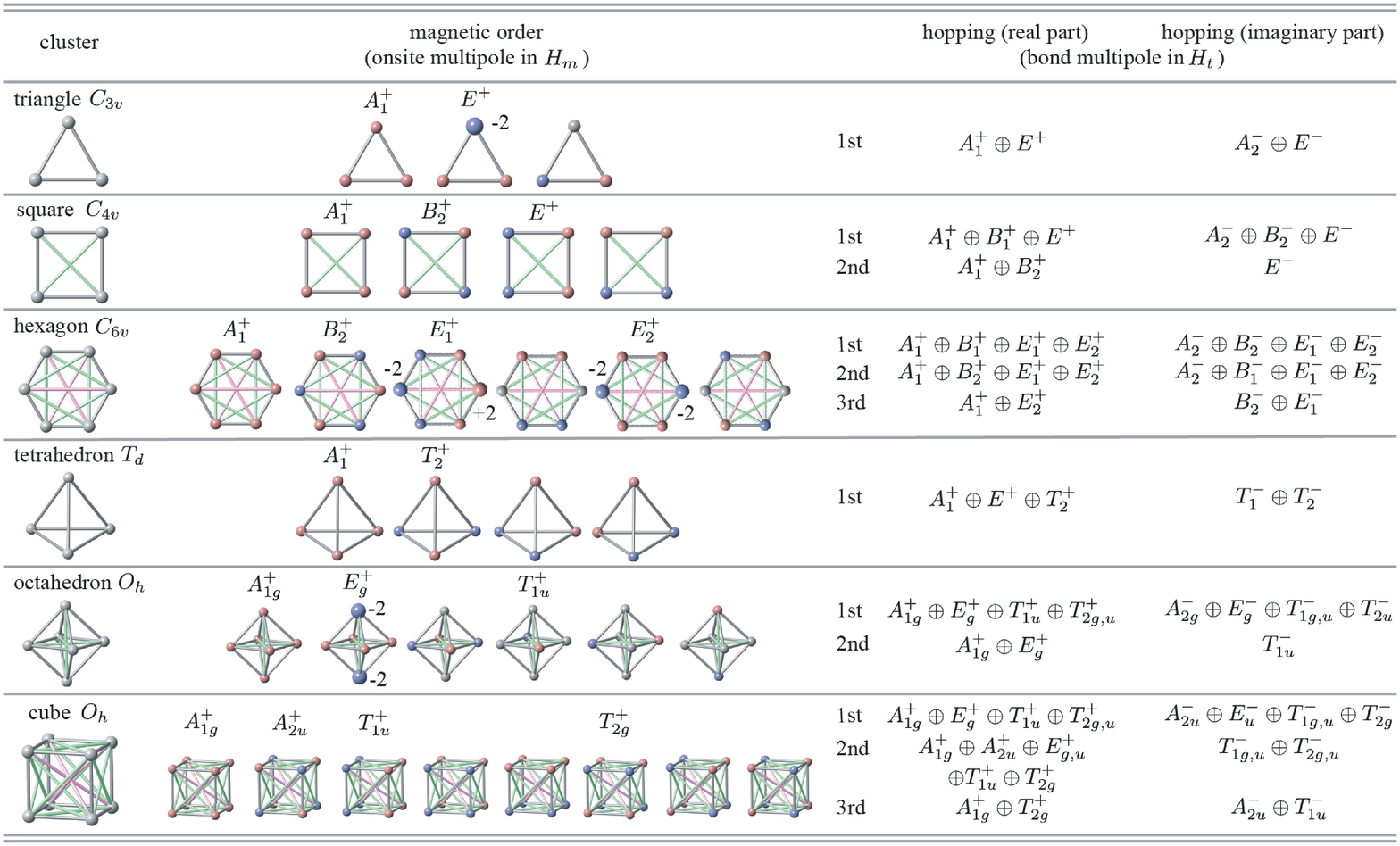} 
\caption{
\label{Fig:cluster}
Irreducible representations of the electric $Q^{(n)}_{\Gamma\gamma}$ and magnetic toroidal $T^{(n)}_{\Gamma\gamma}$multipoles in the MF Hamiltonian, and the 1st-, 2nd-, and 3rd-nearest-neighbor hopping Hamiltonian in fundamental clusters. 
In the column ``magnetic order", the red and blue spheres represent the opposite spin alignment (the gray spheres represent no spin moment) and the size of spheres denotes the amplitude of spins where the number stands for the relative ratio.
The ordered patterns in two- and three-dimensional irreps. can be determined from cluster multipole basis sets~\cite{Suzuki_PhysRevB.99.174407}. 
}
\end{center}
\end{figure*}

\subsection{General classification}
\label{sec:General classification}
Similar analysis can be straightforwardly extended to any other point groups. 
We classify types of second-order spin splitting in $\bm{k}$ according to the irreps. under 32 point groups in Table~\ref{tab_multipoles_table1} (see also Appendix~\ref{sec:Higher-order momentum-dependent spin splittings under 32 point groups} for types of higher-order spin splittings). 
We also show the possible irreps. of the AFM ordering and the electric and magnetic toroidal multipoles of the $n$th-neighbor bond degree of freedom for fundamental clusters.  
Note that the $\bm{k}$-dependent spin splitting under the MF multipole $Q_{\Gamma\gamma}^{(0)}$ 
is also activated by higher-order multiplication of magnetic toroidal multipoles $T_{\nu}^{(n)}$ ($\nu=x,y,z$) in the hopping Hamiltonian. 
In a periodic crystal, the active bond multipoles in the hopping matrix are determined by the translational and site symmetries.

The above multipole decomposition can be also applied to space groups with the glide and/or screw symmetries, i.e., nonsymmorphic space groups, as already shown in the example of pyrochlore lattice having $Fd\bar{3}m$ symmetry in Sec.~\ref{sec:Pyrochlore structure}. 
The organic compound $\kappa$-(BETD-TTF)$_2$Cu[N(CN)$_2$]Cl is a prototype to exhibit the spin splitting in the AFM ordering with the glide symmetry breaking~\cite{naka2019spin}. 
As the two-dimensional AFM pattern belongs to the irrep. $A^{+}_2$ of the point group $C_{2v}$ in Table~\ref{tab_multipoles_table1}, the spin splitting of $Q_{xy}(\bm{k})\sigma \sim k_x k_y \sigma$ appears. 
Microscopically, the 2nd-nearest-neighbor hopping is necessary to obtain such a spin splitting by noting that $A^+_{2}$ under $C_{2v}$ corresponds to $B^+_{2}$ under $C_{4v}$ in Fig.~\ref{Fig:cluster}.

Furthermore, there are good AFM candidate materials to show such a symmetric spin-split band structure~\cite{gallego2016magndata}. 
For example, 
orthorhombic compounds, such as Ca$_3$Mn$_2$O$_7$ (space group \#36 $Cmc2_1$)~\cite{lobanov2004crystal}, 
Cu$_2$V$_2$O$_7$ (space group \#43 $Fdd2$)~\cite{Gitgeatpong_PhysRevB.92.024423}, 
LaMnO$_3$ (space group \#62 $Pnma$)~\cite{Moussa_PhysRevB.54.15149}, 
NaOsO$_3$ (space group \#62 $Pnma$)~\cite{Calder_PhysRevLett.108.257209}, 
and CaIrO$_3$ (space group \#63 $Cmcm$)~\cite{Ohgushi_PhysRevLett.110.217212}, 
tetragonal compounds, such as MnLaMnSbO$_6$ (space group \#86 $P4_2/n$)~\cite{Solana-Madruga_PhysRevB.97.134408}, 
Ba$_2$MnSi$_2$O$_7$ (space group \#113 $P\bar{4}2_1m$)~\cite{sale2019crystal}, 
Ce$_2$Mn$_2$Ge$_4$O$_{12}$ (space group \#125 $P4/nbm$)~\cite{xu2017magnetic}, 
MnF$_2$ (space group \#136 $P4_2/mnm$)~\cite{yamani2010neutron}, 
and CoF$_2$ (space group \#136 $P4_2/mnm$)~\cite{jauch2004gamma}, 
and trigonal compounds, such as Mn$_3$Si$_2$Te$_6$ (space group \#163 $P\bar{3}1c$)~\cite{May_PhysRevB.95.174440}, 
FeBO$_3$ (space group \#167 $R\bar{3}c$)~\cite{Kalashnikova_PhysRevLett.99.167205}, 
and FeCO$_3$ (space group \#167 $R\bar{3}c$)~\cite{jacobs1963metamagnetism}, 
are almost collinear AFM magnets with the small canted and/or weak ferromagnetic moments, which are expected to exhibit the symmetric spin-split band structures, once the bond multipoles are coupled with the AFM onsite multipoles. 
Although the effect of the atomic SOC might also contribute to the spin-splitting in the band structure, especially for the materials with heavier elements, it is important to take into account the contribution induced by the effective spin-orbit interaction between kinetic motions of electrons and the AFM mean field at the quantitative level, since the AFM-driven spin-splitting can be the order of the exchange energy.

It is noteworthy to point out that the anisotropic symmetric spin-split band structure through the effective spin-orbit interaction can be found in noncollinear and noncoplanar magnetic structures by regarding them as a superposition of different spin components. 
For example, the all-in/all-out magnetic structure, which was observed in the pyrochlore compound Cd$_2$Os$_2$O$_7$~\cite{harima2002electronic,Yamaura_PhysRevLett.108.247205}, are expected to exhibit the anisotropic symmetric spin-split band structure in the form of $Q_{yz}(\bm{k})\sigma_x+Q_{zx}(\bm{k})\sigma_y+Q_{xy}(\bm{k})\sigma_z \sim k_y k_z \sigma_z + k_z k_x \sigma_y + k_x k_y \sigma_z$, since the bond multipoles in $H_t$ includes $Q_{yz}(\bm{k})$, $Q_{zx}(\bm{k})$, and $Q_{xy}(\bm{k})$ in Eq.~(\ref{eq:hammmul}). 
Further efforts from both theoretical and experimental sides are highly desired for such exploration.

\section{Physical phenomena by the spin splitting}
\label{sec:Physical phenomena by the spin splitting}
Finally, we discuss characteristic physical phenomena driven by the anisotropic spin splitting. 
One is the spin-current generation in metals by an electric current as 
$J^s_i  = \sigma^{\rm SC}_{ij} J_j$ where $J_i$ and $J^s_i=J_i \sigma$ are electric and spin currents in the $i=x,y,z$ direction ($\sigma$ is in the ordered moment direction) and $\sigma^{\rm SC}$ represents the symmetric spin conductivity tensor, $\sigma^{\rm SC}_{ij}=\sigma^{\rm SC}_{ji}$~\cite{naka2019spin}. 
The components of $\sigma^{\rm SC}$ become nonzero once any of five quadrupoles are active, which are summarized under 32 point groups in Table~\ref{tab_multipoles_table1}. 
For example, the pure spin current along the $x$ ($y$) direction is induced perpendicular to
the electric current along the $y$ ($x$) direction in the cases of $Q_{xy}$ in Figs.~\ref{Fig:spinsplit}(a) and \ref{Fig:spinsplit}(c) in Sec.~\ref{sec: Spin Splittings in tight-binding models}, which leads to a different behavior from the spin-polarized current in the ferromagnets ($Q_0$).
The present spin-current generation is advantageous since the spin current is well-defined quantity without the atomic SOC. 
Likewise, the spin-current is driven by temperature gradient in magnetic insulators, as the electric and thermal currents have the same symmetry property~\cite{naka2019spin}. 

Another interesting response is a magneto-striction (piezomagnetic) effect where a uniform magnetization $M_{i}$ is induced by a strain field $\varepsilon_{jk}$ as $M_{i} = d_{ijk} \varepsilon_{jk}$. 
For example, in the $xy$-type AFM in Figs.~\ref{Fig:spinsplit}(a) and \ref{Fig:spinsplit}(c) in Sec.~\ref{sec: Spin Splittings in tight-binding models}, the uniform magnetization in the ordered-moment direction is induced by applying the shear-type strain field $\varepsilon_{xy}$.

\section{Summary}
\label{sec:Summary}
We investigated a novel route of effective ``spin-orbit" interaction activated by the collinear AFM orderings without the atomic SOC. 
By applying the microscopic multipole descriptions to a tight-binding Hamiltonian, we demonstrated the symmetry conditions and microscopic parameters to obtain such an effective spin-orbit interaction and summarized them for 32 point groups: the multipoles in the hopping matrix or its higher-order multiplication belongs to the same irrep. of the MF multipole in the ordered state. 
We also discussed physical phenomena driven by the spin splitting, such as a spin-current generation and a magneto-striction (piezomagnetic) effect. 
Our microscopic engineering of momentum-dependent spin splitting will encourage for searching further SOC physics even with negligibly small atomic SOCs, such as light-element materials and 3$d$ transition metal oxides. 

\begin{acknowledgments}
The authors are grateful to M. Naka, H. Seo, and Y. Motome for helpful discussions. 
This research was supported by JSPS KAKENHI Grants Numbers JP15H05885, JP18H04296 (J-Physics), JP18K13488, JP19K03752, and JP19H01834. 
This work was also supported by the Toyota Riken Scholarship. 
\end{acknowledgments}

\appendix

\section{Absence of spin splitting in a bipartite system}
\label{sec:Absence of spin splitting in a bipartite system}
We here show that there is no spin-split band structure in a bipartite system due to the chiral symmetry. 
The $2N$-sublattice spinless hopping matrix is generally expressed as 
\begin{align}
\label{eq:Ham_bipar}
H_t=\left(
\begin{array}{cccccc}
 0  & H_{{\bf A}{\bf B}}(\bm{k})   \\
H^\dagger_{{\bf A}{\bf B}}(\bm{k}) &  0   \\
\end{array}
\right), 
\end{align}
where $H_{{\bf A}{\bf B}}(\bm{k})$ is the $N \times N$ matrix. 
As the matrix in Eq.~(\ref{eq:Ham_bipar}) has the chiral symmetry, $H_t$ anticommutes with 
\begin{align}
H_I=\left(
\begin{array}{cccccc}
 I  & 0   \\
0 &  -I   \\
\end{array}
\right), 
\end{align}
where $I$ is the $N \times N$ unit matrix. 
Consequently, the matrix in Eq.~(\ref{eq:Ham_bipar})  has $N$ pairs of eigenvalues as $\pm E_{i}$ ($i=1$-$N$). 

Meanwhile, under the antiferromagnetic collinear ordering where the sublattices ${\bf A}$ and ${\bf B}$ have opposite spin polarizations, the Hamiltonian is expressed as $H_t+ H_{m}$ with $H_{m}=\sigma h  H_I$ for spin $\sigma$. 
As $H_t$ and $H_I$ anticommute with each other, the $N$ pairs of eigenvalues are modified as $\pm \sqrt{E_{i}^2+h^2}$ ($i=1$-$N$).
Therefore, each band are doubly degenerate with respect to the spin $\sigma$ to satisfy $\varepsilon_{\sigma}(\bm{k}) =\varepsilon_{-\sigma}(\bm{k})$.

\section{Effective spin-orbit interaction for different phase conventions}
\label{sec:Effective spin-orbit coupling for different phase conventions}

In Sec.~\ref{sec:Tetragonal structure}, we describe the effective spin-orbit interaction triggered by the collinear antiferromagnetic orderings in sublattice systems. 
To this end, we consider tetragonal crystal structure on the two-dimensional plane in Fig.~2(c) and show how effective spin-orbit interaction through the anisotropic kinetic motions of electrons emerges. 
Although we adopt the phase convention where the phase factor within a unit cell is considered, the obtained results must be irrelevant to choices of phase conventions within a unit cell. 
Here, we show that in the phase convention where the phase factor within a unit cell is {\it not} considered, the higher-order multiplications of the hopping Hamiltonian indeed play the same role as the lowest-order multipoles in the phase convention used in Sec.~\ref{sec:Tetragonal structure}.

The four-sublattice hopping Hamiltonian in the matrix representation is given by 
\begin{widetext}
\begin{align}
\label{eq:Ham_Mat}
H_{t}=\left(
\begin{array}{cccc}
 0 & 
  \begin{array}{c}
 \tilde{Q}_0^{(2)}(\bm{k})+\tilde{Q}_{xy}^{(2)}(\bm{k}) \\+i [\tilde{T}_x^{(2)}(\bm{k})+ \tilde{T}_y^{(2)}(\bm{k})]
 \end{array}
  & \tilde{Q'}_0^{(1)}(\bm{k})-\tilde{Q'}_v^{(1)}(\bm{k})+i \tilde{T}_x^{(1)}(\bm{k}) & \tilde{Q'}_0^{(1)}(\bm{k})+\tilde{Q'}_v^{(1)}(\bm{k})+i \tilde{T}_y^{(1)}(\bm{k}) \\
 {\rm H.c.}& 0 & \tilde{Q'}_0^{(1)}(\bm{k})+\tilde{Q'}_v^{(1)}(\bm{k})-i \tilde{T}_y^{(1)}(\bm{k}) & \tilde{Q'}_0^{(1)}(\bm{k})-\tilde{Q'}_v^{(1)}(\bm{k})-i \tilde{T}_x^{(1)}(\bm{k}) \\
 {\rm H.c.}&  {\rm H.c.} & 0 & 
   \begin{array}{c}
 \tilde{Q}_0^{(2)}(\bm{k})-\tilde{Q}_{xy}^{(2)}(\bm{k}) \\ 
 -i [\tilde{T}_x^{(2)}(\bm{k})- \tilde{T}_y^{(2)}(\bm{k})]
 \end{array} \\
 {\rm H.c.} &  {\rm H.c.} &  {\rm H.c.} & 0 \\
\end{array}
\right),
\end{align}
\end{widetext}
where $\tilde{Q}_{\Gamma\gamma}^{(n)}=Q_{\Gamma\gamma}^{(na)}+Q_{\Gamma\gamma}^{(nb)}$ and $\tilde{Q'}_{\Gamma\gamma}^{(n)}(\bm{k})=\tilde{Q}_{\Gamma\gamma}^{(n)}(\bm{k})/\sqrt{2}$. 
Note that the hopping matrix in Eq.~(\ref{eq:Ham_Mat}) takes account of arbitrary phase conventions including no phase factors within a unit cell, i.e., $a=0$ and $b=1$.
The multipoles in the case of no phase-factor convention is given in Table~\ref{tab_multipoles}. 

We here adopt the basis for the molecular orbitals within 4 sublattice under $C_{4v}$ group, whose functions are given by 
\begin{align}
A_{1}:\quad& \psi_{A_1} = \frac{1}{2} (\psi_{\rm A} + \psi_{\rm B} + \psi_{\rm C} + \psi_{\rm D}), \\
B_{2}:\quad& \psi_{B_2} = \frac{1}{2} (\psi_{\rm A} + \psi_{\rm B} - \psi_{\rm C} - \psi_{\rm D}), \\
E:\quad& \psi_{E^{(1)}} = \frac{1}{2} (-\psi_{\rm A} + \psi_{\rm B} - \psi_{\rm C} + \psi_{\rm D}), \\
& \psi_{E^{(2)}} = \frac{1}{2} (-\psi_{\rm A} + \psi_{\rm B} + \psi_{\rm C} - \psi_{\rm D}), 
\end{align}
where $\psi_{i}$ is the atomic wave function at site $i=$A-D [see Fig.~2(c) in Sec.~\ref{sec:Tetragonal structure}]. 
For these basis functions, the hopping matrix in Eq.~(\ref{eq:Ham_Mat}) is rewritten as 
\begin{widetext}
\begin{align}
\tilde{H}_{t}&=U^{-1}H_{t}U, \\
\label{eq:Ham_MatUni}
&=\left(
\begin{array}{cccc}
2 \tilde{Q'}_0^{(1)}(\bm{k})+ \tilde{Q}_0^{(2)}(\bm{k}) & \tilde{Q}_{xy}^{(2)}(\bm{k}) & i [\tilde{T}_y^{(1)}(\bm{k})+\tilde{T}_y^{(2)}(\bm{k})] & i [\tilde{T}_x^{(1)}(\bm{k})+\tilde{T}_x^{(2)}(\bm{k})] \\
{\rm H.c.} 
& -2 \tilde{Q'}_0^{(1)}(\bm{k})+\tilde{Q}_0^{(2)}(\bm{k}) & -i [\tilde{T}_x^{(1)}(\bm{k})-\tilde{T}_x^{(2)}(\bm{k})] & -i [\tilde{T}_y^{(1)}(\bm{k})-\tilde{T}_y^{(2)}(\bm{k})] \\
{\rm H.c.} 
& {\rm H.c.} 
& -2 \tilde{Q'}_v^{(1)}(\bm{k})-\tilde{Q}_0^{(2)}(\bm{k}) & -\tilde{Q}_{xy}^{(2)}(\bm{k}) \\
{\rm H.c.} 
& {\rm H.c.} 
& {\rm H.c.} 
& 2 \tilde{Q'}_v^{(1)}(\bm{k})-\tilde{Q}_0^{(2)}(\bm{k}) \\
\end{array}
\right),
\end{align}
\end{widetext}
where 
\begin{align}
U=\frac{1}{2}\left(
\begin{array}{cccc}
1 & 1 & -1 & -1 \\
1 & 1 & 1 & 1 \\
1 & -1 & -1 & 1 \\
1 & -1 & 1 & -1
\end{array}
\right).
\end{align}

Regarding the off-diagonal element formally as a perturbation, we show that the effective electric quadrupole $\tilde{Q}^{(0)}_{xy}$ comes out, which couples with the mean-field $h_{xy}$. 
For the basis functions $\psi_{A_1}$ and $\psi_{B_2}$, $\tilde{Q}^{(0)}_{xy}$ is expressed up to the second order in $\bm{k}$ by 
\begin{widetext}
\begin{align}
\label{eq:Qxy_A1}
\braket{\psi_{A_1}|U^{-1} Q^{(0)}_{xy} U|\psi_{A_1}}=
&\frac{\tilde{Q}_{xy}^{(2)}(\bm{k})}{2 \tilde{Q'}_0^{(1)}(\bm{k})}
+
\frac{ [2 \tilde{T}_x^{(2)}(\bm{k}) \tilde{T}_y^{(2)}(\bm{k})+\tilde{T}_x^{(1)}(\bm{k}) \tilde{T}_y^{(2)}(\bm{k})+ \tilde{T}_y^{(1)}(\bm{k})\tilde{T}_x^{(2)}(\bm{k})]}{2  [\tilde{Q'}_0^{(1)}(\bm{k})+\tilde{Q}_0^{(2)}(\bm{k})]^2}
\nonumber \\
&-
\frac{\tilde{Q}_0^{(2)}(\bm{k}) [\tilde{T}_x^{(1)}(\bm{k}) \tilde{T}_y^{(1)}(\bm{k})-\tilde{T}_x^{(2)}(\bm{k}) \tilde{T}_y^{(2)}(\bm{k})]}{2 \tilde{Q'}_0^{(1)}(\bm{k}) [\tilde{Q'}_0^{(1)}(\bm{k})+\tilde{Q}_0^{(2)}(\bm{k})]^2}+\mathcal{O}(k^3),\\
\label{eq:Qxy_B2}
\braket{\psi_{B_2}| U^{-1} Q^{(0)}_{xy} U|\psi_{B_2}}=
&-\frac{\tilde{Q}_{xy}^{(2)}(\bm{k})}{2 \tilde{Q'}_0^{(1)}(\bm{k})}
+
\frac{ [2 \tilde{T}_x^{(2)}(\bm{k}) \tilde{T}_y^{(2)}(\bm{k})-\tilde{T}_x^{(1)}(\bm{k}) \tilde{T}_y^{(2)}(\bm{k})- \tilde{T}_y^{(1)}(\bm{k})\tilde{T}_x^{(2)}(\bm{k})]}{2  [\tilde{Q'}_0^{(1)}(\bm{k})-\tilde{Q}_0^{(2)}(\bm{k})]^2}
\nonumber \\
&+
\frac{\tilde{Q}_0^{(2)}(\bm{k}) [\tilde{T}_x^{(1)}(\bm{k}) \tilde{T}_y^{(1)}(\bm{k})-\tilde{T}_x^{(2)}(\bm{k}) \tilde{T}_y^{(2)}(\bm{k})]}{2 \tilde{Q'}_0^{(1)}(\bm{k}) [\tilde{Q'}_0^{(1)}(\bm{k})-\tilde{Q}_0^{(2)}(\bm{k})]^2}+\mathcal{O}(k^3). 
\end{align}
\end{widetext}
The results in Eqs.~(\ref{eq:Qxy_A1}) and (\ref{eq:Qxy_B2}) clearly indicate that the hopping matrix $\tilde{Q}^{(2)}_{xy}(\bm{k})$ or its higher-order multiplication, $\tilde{T}^{(n)}_x (\bm{k}) \tilde{T}^{(m)}_y (\bm{k})$ belongs to the same irrep. of $B_2$ of the mean-field multipole $Q^{(0)}_{xy}$ in the ordered state. 
Moreover, it is shown that there is no spin-split band structure in the absence of multipoles $\tilde{Q}^{(2)}_{\Gamma\gamma}$ and $\tilde{T}^{(2)}_{\Gamma\gamma}$, i.e., the absence of the next-nearest-neighbor hopping, as shown by the dashed lines in Fig.~2(d). 

\begin{table*}[t!]
\caption{
Multipoles in two-dimensional tetragonal structure classified by $C_{4v}$.
The phase factor within a unit cell is not considered. 
We use the abbreviations, $c_{\mu}^{\eta}=\cos k_{\mu}\eta$, $s_{\mu}^{\eta}=\sin k_{\mu}\eta$
for $\mu=x, y$ and  $\eta=a,b$.
$\gamma_{\pm}=(\rho_{x}\tau_{x}\pm\tau_{x})/\sqrt{2}$.
\\
}
\label{tab_multipoles}
\centering
\begin{tabular}{cccccccc} \hline\hline
irrep. & type & $Q_{\Gamma\gamma}^{(1)}$ & $Q_{\Gamma\gamma}^{(2)}$ & $Q_{\Gamma\gamma}^{(1a)}(\bm{k})$ & $Q_{\Gamma\gamma}^{(1b)}(\bm{k})$ & $Q_{\Gamma\gamma}^{(2a)}(\bm{k})$ & $Q_{\Gamma\gamma}^{(2b)}(\bm{k})$ \\
& & $T_{\Gamma\gamma}^{(1)}$ & $T_{\Gamma\gamma}^{(2)}$ & $T_{\Gamma\gamma}^{(1a)}(\bm{k})$ & $T_{\Gamma\gamma}^{(1b)}(\bm{k})$ & $T_{\Gamma\gamma}^{(2a)}(\bm{k})$ & $T_{\Gamma\gamma}^{(2b)}(\bm{k})$\\ \hline
$A_{1}^{+}$ & $Q_{0}$ & $\gamma_{+}$ & $\rho_{x}$ & $\sqrt{2} t^{(1)}_{a}$ & $\frac{1}{\sqrt{2}}(c_{x}^{b}+c_{y}^{b})t^{(1)}_{b}$ & $t^{(2)}_{a}$ & $c_{x}^{b}c_{y}^{b}t^{(2)}_{b}$ \\ \hline
$B_{1}^{+}$ & $Q_{v}$ & $\gamma_{-}$ & &  & $-\frac{1}{\sqrt{2}}(c_{x}^{b}-c_{y}^{b})t^{(1)}_{b}$ & & \\ \hline
$B_{2}^{+}$ & $Q_{xy}$  & & $\rho_{x}\tau_{z}$ & & &  &$-s_{x}^{b}s_{y}^{b}t^{(2)}_{b}$ \\ \hline
$E^{+}$ & $Q_{yz}$ & $-\rho_z \tau_x$ & & & & & \\
& $Q_{zx}$ & $\rho_y \tau_y$ & & & & & \\ \hline
$E^{-}$ & $T_{x}$ & $-\rho_{z}\tau_{y}$ & $-\rho_{y}\tau_{z}$ &  & $-s_{x}^{b}t^{(1)}_{b}$ &  & $-s_{x}^{b}c_{y}^{b}t^{(2)}_{b}$ \\
& $T_{y}$ & $-\rho_{y}\tau_{x}$ & $-\rho_{y}$ &  & $-s_{y}^{b}t^{(1)}_{b}$ &  & $-s_{y}^{b}c_{x}^{b}t^{(2)}_{b}$ \\ \hline\hline
\end{tabular}
\end{table*}

By substituting $\tilde{Q'}_0^{(1)}(\bm{k}) =  t^{(1)}_a+t^{(1)}_b, \tilde{Q}_0^{(2)}(\bm{k}) =  t^{(2)}_a+t^{(2)}_b, \tilde{Q}_{xy}^{(2)}(\bm{k}) = -a^2  k_x  k_y t^{(2)}_a -  b^2 k_x k_y t^{(2)}_b, \tilde{T}_x^{(n)}(\bm{k}) = a k_x t^{(n)}_a-b k_x t^{(n)}_b, \tilde{T}_y^{(n)}(\bm{k}) = a k_y t^{(n)}_a-b k_y t^{(n)}_b$, which are obtained from the multipole expressions in Table II in the limit of $\bm{k} \to 0$, Eqs.~(\ref{eq:Qxy_A1}) and (\ref{eq:Qxy_B2}) read 
\begin{widetext}
\begin{align}
\label{eq:Qxy_A1_t}
&\braket{\psi_{A_1}|U^{-1} Q^{(0)}_{xy} U|\psi_{A_1}}=
-\frac{  \left[t^{(2)}_b (t^{(1)}_a+t^{(2)}_a)^2+t^{(2)}_a (t^{(1)}_b+t^{(2)}_b)^2\right]}{2 (t^{(1)}_a+t^{(1)}_b) (t^{(1)}_a+t^{(1)}_b+t^{(2)}_a+t^{(2)}_b)^2}(a+b)^2 k_x k_y+\mathcal{O}(k^3),\\
\label{eq:Qxy_B2_t}
&\braket{\psi_{B_2}|U^{-1} Q^{(0)}_{xy} U|\psi_{B_2}}=
\frac{  \left[t^{(2)}_b (t^{(1)}_a-t^{(2)}_a)^2+t^{(2)}_a (t^{(1)}_b-t^{(2)}_b)^2\right]}{2 (t^{(1)}_a+t^{(1)}_b) (t^{(1)}_a+t^{(1)}_b-t^{(2)}_a-t^{(2)}_b)^2}(a+b)^2 k_x k_y+\mathcal{O}(k^3). 
\end{align}
\end{widetext}
The factor $(a+b)^2$ implies that a choice of the phase factor is not important. 
Note that the expressions in Eqs.~(\ref{eq:Qxy_A1_t}) and (\ref{eq:Qxy_B2_t}) include the case of no phase-factor convention by setting $a=0$ and $b=1$. 
In fact, the same expressions are obtained by substituting the multipole expressions in Table~\ref{tab_multipoles} into Eqs.~(\ref{eq:Qxy_A1}) and (\ref{eq:Qxy_B2}).

\section{Higher-order momentum-dependent spin splittings under 32 point groups}
\label{sec:Higher-order momentum-dependent spin splittings under 32 point groups}
We summarize types of spin-split band structures up to the sixth order in $\bm{k}$ according to the irreps. under 32 crystallographic point groups in Tables~\ref{tab_multipoles_table1} and \ref{tab_multipoles_table2}, as discussed in Sec.~\ref{sec:General classification}. 

\begin{table*}[t!]
\caption{
Multipoles (MP), rank, and spin splitting (SS) under cubic, tetragonal, orthorhombic, monoclinic, and triclinic crystals. 
We take the $x$ axis as the $C_{2}'$ rotation axis. 
The unlisted point groups are as follows: $T_{d}$ is the same as $O$. 
$D_{2d}$ and $C_{4v}$ are the same as $D_4$. 
$S_{4}$ is the same as $C_4$. 
}
\label{tab_multipoles_table1}
\centering
\begin{tabular}{c|c|c|cc|cc|cc|cc|c} \hline\hline
MP & rank & SS & 
$O_{(h)}$ & $T_{(h)}$ &
$D_{4(h)}$ & $C_{4(h)}$ & 
$D_{2(h)}$ & $C_{2v}$ & $C_{2(h)}$ & $C_{s}$ &
$C_{(i)}$ 
\\ \hline
$Q^{(0)}_{0}$ & 0 & $\sigma$ & 
$A^+_{1(g)}$ & $A^+_{(g)}$ &
$A^+_{1(g)}$ & $A^+_{(g)}$ & 
$A^+_{(g)}$ & $A^+_{1}$ & $A^+_{(g)}$ & $A'^+$ 
& 
$A^+_{(g)}$ 
\\ \hline
$Q^{(0)}_{u}$ & 2 & $(3k_z^2-k^2)\sigma$ & 
$E^+_{(g)}$ & $E^+_{(g)}$ & 
$A^+_{1(g)}$ & $A^+_{(g)}$ & 
$A^+_{(g)}$ & $A^+_{1}$ & $A^+_{(g)}$ & $A'^+$ 
& 
$A^+_{(g)}$ 
\\
$Q^{(0)}_{v}$ & & $(k_x^2-k_y^2)\sigma$ & 
& &
$B^+_{1(g)}$ & $B^+_{(g)}$ & 
$A^+_{(g)}$ & $A^+_{1}$ & $A^+_{(g)}$ & $A'^+$ 
& 
$A^+_{(g)}$ 
\\
$Q^{(0)}_{xy}$ &  & $k_x k_y\sigma$ & 
$T^+_{2(g)}$ & $T^+_{(g)}$ & 
$B^+_{2(g)}$ & $B^+_{(g)}$ & 
$B^+_{1(g)}$ & $A^+_{2}$ & $A^+_{(g)}$ & $A'^+$ 
&
$A^+_{(g)}$ 
\\ 
$Q^{(0)}_{yz}$ & & $k_y k_z\sigma$ & 
& &
$E^+_{(g)}$ & $E^+_{(g)}$ & 
$B^+_{3(g)}$ & $B^+_{2}$ & $B^+_{(g)}$ & $A''^+$ 
& 
$A^+_{(g)}$ 
\\
$Q^{(0)}_{zx}$ & & $k_z k_x \sigma$ & 
& &
& & 
$B^+_{2(g)}$ & $B^+_{1}$ & $B^+_{(g)}$ & $A''^+$ 
&
$A^+_{(g)}$ 
\\ \hline
$Q^{(0)}_{4}$ & 4 &$(k_x^4+k_y^4+k_z^4)\sigma$ & 
$A^+_{1(g)}$ & $A^+_{(g)}$ &
$A^+_{1(g)}$ & $A^+_{(g)}$ & 
$A^+_{(g)}$ & $A^+_{1}$ & $A^+_{(g)}$ & $A'^+$ 
& 
$A^+_{(g)}$ 
\\
$Q^{(0)}_{4u}$ & & $(k_x^4-k_y^4)\sigma$ & 
$E^+_{(g)}$ & $E^+_{(g)}$ & 
$A^+_{1(g)}$ & $A^+_{(g)}$ & 
$A^+_{(g)}$ & $A^+_{1}$ & $A^+_{(g)}$ & $A'^+$ 
& 
$A^+_{(g)}$ 
\\
$Q^{(0)}_{4v}$ & & $(2k_z^4-k_x^4-k_y^4)\sigma$ & 
& &
$B^+_{1(g)}$ & $B^+_{(g)}$ & 
$A^+_{(g)}$ & $A^+_{1}$ & $A^+_{(g)}$ & $A'^+$ 
& 
$A^+_{(g)}$ 
\\
$Q^{\alpha (0)}_{4z}$ & & $k_x k_y (k_x^2-k_y^2) \sigma$ & 
$T^+_{1(g)}$ & $T^+_{(g)}$ &
$A^+_{2(g)}$ & $A^+_{(g)}$ & 
$B^+_{1(g)}$ & $A^+_{2}$ & $A^+_{(g)}$ & $A'^+$
&
$A^+_{(g)}$ 
\\
$Q^{\alpha (0)}_{4x}$ & & $k_y k_z (k_y^2-k_z^2) \sigma$ & 
& &
$E^+_{(g)}$ & $E^+_{(g)}$ & 
$B^+_{3(g)}$ & $B^+_{2}$ & $B^+_{(g)}$ & $A''^+$ 
& 
$A^+_{(g)}$ 
\\
$Q^{\alpha (0)}_{4y}$ & & $k_z k_x (k_z^2-k_x^2) \sigma$ & 
& &
& & 
$B^+_{2(g)}$ & $B^+_{1}$ & $B^+_{(g)}$ & $A''^+$ 
&
$A^+_{(g)}$ 
\\
$Q^{\beta (0)}_{4z}$ & & $k_z^2 k_x k_y \sigma$ & 
$T^+_{2(g)}$ & $T^+_{(g)}$ & 
$B^+_{2(g)}$ & $B^+_{(g)}$ & 
$B^+_{1(g)}$ & $A^+_{2}$ & $A^+_{(g)}$ & $A'^+$ 
&
$A^+_{(g)}$ 
\\
$Q^{\beta (0)}_{4x}$ &  & $k_x^2 k_y k_z \sigma$ & 
& &
$E^+_{(g)}$ & $E^+_{(g)}$ & 
$B^+_{3(g)}$ & $B^+_{2}$ & $B^+_{(g)}$ & $A''^+$ 
& 
$A^+_{(g)}$ 
\\
$Q^{\beta (0)}_{4y}$ & & $k_y^2 k_z k_x \sigma$ & 
& &
& & 
$B^+_{2(g)}$ & $B^+_{1}$ & $B^+_{(g)}$ & $A''^+$ 
&
$A^+_{(g)}$ 
\\ \hline
$Q^{(0)}_{6}$ & 6 & $k_x^2 k_y^2 k_z^2 \sigma$ & 
$A^+_{1(g)}$ & $A^+_{(g)}$ &
$A^+_{1(g)}$ & $A^+_{(g)}$ & 
$A^+_{(g)}$ & $A^+_{1}$ & $A^+_{(g)}$ & $A'^+$ 
& 
$A^+_{(g)}$ 
\\
$Q^{(0)}_{6t}$ & & $[k_x^4 (k_y^2 - k_z^2)+k_y^4 (k_z^2 - k_x^2)+k_z^4 (k_x^2 - k_y^2)] \sigma$ & 
$A^+_{2(g)}$ & $A^+_{(g)}$ &
$B^+_{1(g)}$ & $B^+_{(g)}$ & 
$A^+_{(g)}$ & $A^+_{1}$ & $A^+_{(g)}$ & $A'^+$ 
& 
$A^+_{(g)}$ 
\\
$Q^{(0)}_{6u}$ & & $(2k_z^6-k_x^6-k_y^6) \sigma$ & 
$E^+_{(g)}$ & $E^+_{(g)}$ & 
$A^+_{1(g)}$ & $A^+_{(g)}$ & 
$A^+_{(g)}$ & $A^+_{1}$ & $A^+_{(g)}$ & $A'^+$ 
& 
$A^+_{(g)}$ 
\\
$Q^{(0)}_{6v}$ & & $(k_x^6-k_y^6)\sigma$ & 
& &
$B^+_{1(g)}$ & $B^+_{(g)}$ & 
$A^+_{(g)}$ & $A^+_{1}$ & $A^+_{(g)}$ & $A'^+$ 
& 
$A^+_{(g)}$ 
\\
$Q^{\alpha (0)}_{6z}$ & & $k_x k_y (k_x^4-k_y^4)\sigma$ & 
$T^+_{1(g)}$ & $T^+_{(g)}$ &
$A^+_{2(g)}$ & $A^+_{(g)}$ & 
$B^+_{1(g)}$ & $A^+_{2}$ & $A^+_{(g)}$ & $A'^+$
&
$A^+_{(g)}$ 
\\
$Q^{\alpha (0)}_{6x}$ & & $k_y k_z (k_y^4-k_z^4)\sigma$ & 
& &
$E^+_{(g)}$ & $E^+_{(g)}$ & 
$B^+_{3(g)}$ & $B^+_{2}$ & $B^+_{(g)}$ & $A''^+$ 
& 
$A^+_{(g)}$ 
\\
$Q^{\alpha (0)}_{6y}$ & & $k_z k_x (k_z^4-k_x^4)\sigma$ & 
& &
& & 
$B^+_{2(g)}$ & $B^+_{1}$ & $B^+_{(g)}$ & $A''^+$ 
&
$A^+_{(g)}$ 
\\
$Q^{\beta1 (0)}_{6z}$ & & $k_x^3 k_y^3\sigma$ & 
$T^+_{2(g)}$ & $T^+_{(g)}$ & 
$B^+_{2(g)}$ & $B^+_{(g)}$ & 
$B^+_{1(g)}$ & $A^+_{2}$ & $A^+_{(g)}$ & $A'^+$ 
&
$A^+_{(g)}$ 
\\
$Q^{\beta1 (0)}_{6x}$ & & $k_y^3 k_z^3\sigma$ & 
& &
$E^+_{(g)}$ & $E^+_{(g)}$ & 
$B^+_{3(g)}$ & $B^+_{2}$ & $B^+_{(g)}$ & $A''^+$ 
& 
$A^+_{(g)}$ 
\\
$Q^{\beta1 (0)}_{6y}$ & & $k_z^3 k_x^3\sigma$ & 
& &
& & 
$B^+_{2(g)}$ & $B^+_{1}$ & $B^+_{(g)}$ & $A''^+$ 
&
$A^+_{(g)}$ 
\\
$Q^{\beta2 (0)}_{6z}$ & & $k_z^4 k_x k_y\sigma$ & 
$T^+_{2(g)}$ & $T^+_{(g)}$ & 
$B^+_{2(g)}$ & $B^+_{(g)}$ & 
$B^+_{1(g)}$ & $A^+_{2}$ & $A^+_{(g)}$ & $A'^+$ 
&
$A^+_{(g)}$ 
\\
$Q^{\beta2 (0)}_{6x}$ & & $k_x^4 k_y k_z\sigma$ & 
& &
$E^+_{(g)}$ & $E^+_{(g)}$ & 
$B^+_{3(g)}$ & $B^+_{2}$ & $B^+_{(g)}$ & $A''^+$ 
& 
$A^+_{(g)}$ 
\\
$Q^{\beta2 (0)}_{6y}$ & & $k_y^4 k_z k_x\sigma$ & 
& &
& & 
$B^+_{2(g)}$ & $B^+_{1}$ & $B^+_{(g)}$ & $A''^+$ 
&
$A^+_{(g)}$ 
\\
\hline\hline
\end{tabular}
\end{table*}

\begin{table*}[t!]
\caption{
Multipoles (MP), rank, and spin splitting (SS) under hexagonal and trigonal crystals. 
We take the $zx$ plane as the $\sigma_v$ mirror plane for $C_{3 {v}}$. 
}
\label{tab_multipoles_table2}
\centering
\begin{tabular}{c|c|c|ccccc|ccc} \hline\hline
MP & rank & SS & 
$D_{6(h)}$ & $C_{6(h)}$ & $C_{6v}$ & $D_{3h}$ & $C_{3h}$ &
$D_{3(d)}$ & $C_{3v}$  & $C_{3(i)}$ 
\\ \hline
$Q^{(0)}_{0}$ & 0 & $\sigma$  & 
$A^+_{1(g)}$ & $A^+_{(g)}$ & $A^+_{1}$ & $A'^+_1$ & $A'^+$ &
$A^+_{1(g)}$ & $A^+_{1}$ & $A^+_{(g)}$ 
\\ \hline
$Q^{(0)}_{u}$ & 2 & $(3k_z^2-k^2)\sigma$ & 
$A^+_{1(g)}$ & $A^+_{(g)}$ & $A^+_{1}$ & $A'^+_1$ & $A'^+$ &
$A^+_{1(g)}$ & $A^+_{1}$ & $A^+_{(g)}$  
\\
$Q^{(0)}_{v}$ & & $(k_x^2-k_y^2)\sigma$ & 
$E^+_{2(g)}$ & $E^+_{2(g)}$ & $E^+_{2}$ & $E'^+$ & $E'^+$ &
$E^+_{(g)}$ & $E^+$ & $E^+_{(g)}$  
\\
$Q^{(0)}_{xy}$ & & $k_x k_y\sigma$ & 
& & & & &
 &  &
\\ 
$Q^{(0)}_{yz}$ & & $k_y k_z\sigma$ & 
$E^+_{1(g)}$ & $E^+_{1(g)}$ & $E^+_{1}$ & $E''^+$ & $E''^+$ &
$E^+_{(g)}$ & $E^+$ & $E^+_{(g)}$  
\\
$Q^{(0)}_{zx}$ & & $k_z k_x \sigma$ & 
& & & & &
 &  & 
\\ \hline
$Q^{(0)}_{40}$ & 4 &$k_z^4\sigma$ & 
$A^+_{1(g)}$ & $A^+_{(g)}$ & $A^+_{1}$ & $A'^+_1$ & $A'^+$ &
$A^+_{1(g)}$ & $A^+_{1}$ & $A^+_{(g)}$ 
\\
$Q^{(0)}_{4a}$ & & $k_y k_z (3k_x^2-k_y^2)\sigma$ & 
$B^+_{1(g)}$ & $B^+_{(g)}$ & $B^+_{2}$ & $A''^+_1$ & $A''^+$ &
$A^+_{1(g)}$ & $A^+_{2}$ & $A^+_{(g)}$
\\
$Q^{(0)}_{4b}$ & & $k_z k_x (k_x^2-3 k_y^2)\sigma$ & 
$B^+_{2(g)}$ & $B^+_{(g)}$ & $B^+_{1}$ & $A''^+_2$ & $A''^+$ &
$A^+_{2(g)}$ & $A^+_{1}$  & $A^+_{(g)}$ 
\\
$Q^{\alpha (0)}_{4u}$ & & $k_x k_z^3 \sigma$ & 
$E^+_{1(g)}$ & $E^+_{1(g)}$ & $E^+_{1}$ & $E''^+$ & $E''^+$ &
$E^+_{(g)}$ & $E^+$ & $E^+_{(g)}$  
\\
$Q^{\alpha (0)}_{4v}$ & & $k_y k_z^3 \sigma$ & 
& & & & &
 &  & 
\\
$Q^{\beta1 (0)}_{4u}$ & & $(k_x^4 - 6k_x^2 k_y^2 + k_y^4) \sigma$ & 
$E^+_{2(g)}$ & $E^+_{2(g)}$ & $E^+_{2}$ & $E'^+$ & $E'^+$ &
$E^+_{(g)}$ & $E^+$ & $E^+_{(g)}$  
\\
$Q^{\beta1 (0)}_{4v}$ & & $k_x k_y (k_x^2-k_y^2)  \sigma$ & 
& & & & &
 &  & 
\\
$Q^{\beta2 (0)}_{4u}$ &  & $k_z^2 (k_x^2-k_y^2)   \sigma$ & 
$E^+_{2(g)}$ & $E^+_{2(g)}$ & $E^+_{2}$ & $E'^+$ & $E'^+$ &
$E^+_{(g)}$ & $E^+$ & $E^+_{(g)}$  
\\
$Q^{\beta2 (0)}_{4v}$ & & $k_z^2 k_x k_y \sigma$ & 
& & & & &
 &  & 
\\ \hline
$Q^{(0)}_{60}$ & 6 & $k_z^6 \sigma$ & 
$A^+_{1(g)}$ & $A^+_{(g)}$ & $A^+_{1}$ & $A'^+_1$ & $A'^+$ &
$A^+_{1(g)}$ & $A^+_{1}$ & $A^+_{(g)}$ 
\\
$Q^{(0)}_{6c}$ & & $[k_x^2 (k_x^2-3k_y^2)^2-k_y^2 (3k_x^2-k_y^2)^2 ]\sigma$ & 
$A^+_{1(g)}$ & $A^+_{(g)}$ & $A^+_{1}$ & $A'^+_1$ & $A'^+$ &
$A^+_{1(g)}$ & $A^+_{1}$ & $A^+_{(g)}$ 
\\
$Q^{(0)}_{6s}$ & & $k_x k_y (k_x^2-3k_y^2) (3k_x^2-k_y^2) \sigma$ & 
$A^+_{2(g)}$ & $A^+_{(g)}$ & $A^+_{2}$ & $A'^+_2$ & $A'^+$ &
$A^+_{2(g)}$ & $A^+_{2}$ & $A^+_{(g)}$ 
\\
$Q^{(0)}_{6a}$ & & $k_y k_z^3 (3k_x^2-k_y^2)\sigma$ & 
$B^+_{1(g)}$ & $B^+_{(g)}$ & $B^+_{2}$ & $A''^+_1$ & $A''^+$ &
$A^+_{1(g)}$ & $A^+_{2}$ & $A^+_{(g)}$
\\
$Q^{(0)}_{6b}$ & & $k_x k_z^3 (k_x^2-3k_y^2)\sigma$ & 
$B^+_{2(g)}$ & $B^+_{(g)}$ & $B^+_{1}$ & $A''^+_2$ & $A''^+$ &
$A^+_{2(g)}$ & $A^+_{1}$  & $A^+_{(g)}$ 
\\
$Q^{\alpha1 (0)}_{6u}$ & & $k_z k_x (k_x^4 - 10 k_x^2 k_y^2 +5 k_y^4)\sigma$ & 
$E^+_{1(g)}$ & $E^+_{1(g)}$ & $E^+_{1}$ & $E''^+$ & $E''^+$ &
$E^+_{(g)}$ & $E^+$ & $E^+_{(g)}$  
\\
$Q^{\alpha1 (0)}_{6v}$ & & $k_y k_z (5 k_x^4 - 10 k_x^2 k_y^2 + k_y^4) \sigma$ & 
& & & & &
 &  & 
\\
$Q^{\alpha2 (0)}_{6u}$ & & $k_x k_z^5\sigma$ & 
$E^+_{1(g)}$ & $E^+_{1(g)}$ & $E^+_{1}$ & $E''^+$ & $E''^+$ &
$E^+_{(g)}$ & $E^+$ & $E^+_{(g)}$  
\\
$Q^{\alpha2 (0)}_{6v}$ & & $k_y k_z^5\sigma$ & 
& & & & &
 &  & 
\\
$Q^{\beta1 (0)}_{6u}$ & & $(k_x^4 - 6k_x^2 k_y^2 + k_y^4) k_z^2  \sigma$ & 
$E^+_{2(g)}$ & $E^+_{2(g)}$ & $E^+_{2}$ & $E'^+$ & $E'^+$ &
$E^+_{(g)}$ & $E^+$ & $E^+_{(g)}$  
\\
$Q^{\beta1 (0)}_{6v}$ & & $k_x k_y (k_x^2-k_y^2)k_z^2 \sigma$ & 
& & & & &
 &  & 
\\
$Q^{\beta2 (0)}_{6u}$ & & $(k_x^2- k_y^2) k_z^4 \sigma$ & 
$E^+_{2(g)}$ & $E^+_{2(g)}$ & $E^+_{2}$ & $E'^+$ & $E'^+$ &
$E^+_{(g)}$ & $E^+$ & $E^+_{(g)}$  
\\
$Q^{\beta2 (0)}_{6v}$ & & $k_x k_y k_z^4\sigma$ & 
& & & & &
 &  & 
\\
\hline\hline
\end{tabular}
\end{table*}

\bibliographystyle{apsrev}
\bibliography{ref}

\begin{thebibliography}{54}
\bibitem[{com({\natexlab{a}})}]{comment_SS2}
\bibinfo{note}{The classification of the spin-split band structure in the
  presence of the SOC is shown in Ref.~\cite{Hayami_PhysRevB.98.165110}.}
\bibitem[{com({\natexlab{b}})}]{comment_phase}
\bibinfo{note}{The phase factor reflecting each sublattice position directly
  represents the symmetry within a unit cell, although the phase convention is
  arbitrary. See also Appendix~\ref{sec:Effective spin-orbit coupling for
  different phase conventions}.}
\end{thebibliography}

\end{document}